 \documentstyle[prd,preprint,aps]{revtex}

\newcommand{\be}{\begin{equation}}
\newcommand{\ee}{\end{equation}}
\newcommand{\bea}{\begin{eqnarray}\maketitle}
\newcommand{\eea}{\end{eqnarray}}

\begin{document}

\reversemarginpar
\tighten

\title{Quasinormal modes for single horizon black holes
 in generic 2-d dilaton gravity} 

\author{Joanne Kettner~$^1$, Gabor Kunstatter~$^1$ and A.J.M. Medved~$^2$}

\address{
$^1$~Physics Department\\
University of Winnipeg \\
Winnipeg, Canada R3B-2E9 \\ 
E-mail: g.kunstatter@uwinnipeg.ca \\ 
and\\
 $^2$~School of Mathematical and Computing Sciences\\
Victoria University of Wellington\\
PO Box 600, Wellington, New Zealand \\
E-Mail: joey.medved@mcs.vuw.ac.nz}

\maketitle

\begin{abstract}

There has been some recent speculation that 
a connection may exist between the quasinormal-mode spectra
of highly damped black holes and the fundamental 
theory of quantum gravity. This notion follows from
a conjecture by Hod that the real part of the highly damped mode frequencies
can be used to calibrate the semi-classical level spacing in the black
hole quantum area spectrum.  However, even if  the level spacing can
be fixed in this manner, it still remains unclear
whether this implies a physically significant  ``duality'' or  
merely a numerical coincidence.  This tapestry of ideas  
serves as the motivation for the current paper. We
utilize the ``monodromy approach''  to calculate  the quasinormal-mode 
spectra for  a generic class
of black holes in two-dimensional dilatonic gravity.
Our results agree with the prior literature whenever a direct
 comparison is
possible and provide the analysis of  a much more
diverse class of black hole models than previously considered. 

\end{abstract}

\section{Introduction}

There has been a longstanding belief  that the fundamental theory of
 quantum gravity --- whatever that might be --- should be 
able to resolve the
theoretical problems of black hole physics.   Most notoriously,
 do black holes
lose information \cite{Hawking-1}, what are the microstates 
responsible for
the (Bekenstein--Hawking) black hole entropy \cite{Bekenstein-1,Hawking-2}, 
and should we worry  about the apparent ``trans-Planckian'' effects
of Hawking radiation \cite{Hawking-2}?
(For some recent discussion on these issues, see, respectively,
\cite{Giddings,Fursaev,Helfer}.) Turned around, it seems reasonable to
suggest that
black holes might be able to tell us something about quantum gravity.
Indeed, it has become popular to look
for ``imprints''  from the fundamental theory at the level of classical
(or semi-classical)
black hole physics. 
For example, it has been argued  ---  on the basis  of black 
hole thought experiments ---  that there should be a  {\it fundamental} limit
 on  information storage
capacity; with this limit coming in  the guise of various  entropy bounds 
\cite{Bekenstein-2,tHooft,Susskind}.

A similar idea, philosophically speaking, is the ``Hod conjecture'' 
\cite{Hod-1},
which relates  certain {\it quantum}  (spectral) features of a black hole to  
the quasinormal
modes of its {\it classical} perturbations.
In particular, Hod has suggested that the  highly damped quasinormal
modes correspond to fundamental vibrational modes of a
black hole horizon. As such, the Bohr correspondence principle
then implies that the vibrational frequencies associated
with these modes must appear as transition frequencies in
the semi-classical energy spectrum of a quantized black hole.
As this conjecture
is a central motivation for the current work, let us now elaborate
on the conceptual framework. 

First of all,  the term ``quasinormal modes'' means, in a general
sense,  a discrete  set of 
{\it complex}-frequency
modes that  are associated with any damped physical system 
and can be identified with complex poles in the scattering amplitude
for the system under consideration. For a black hole system
in particular,~\footnote{For an
comprehensive review article on the quasinormal modes of black holes,
see \cite{Kokkotas}.}     small perturbations arising near the horizon will
give rise to  quasinormal modes on account of the 
spacetime  curvature acting like a damping potential.
Moreover, these quasinormal modes are expected to dominate
the gravitational wave signal associated with 
the dynamics of astrophysical black holes.

Given a black  hole in an asymptotically flat spacetime, 
it has been demonstrated 
that any perturbation will generate 
an infinite set of such modes \cite{Bachelot} with a spectrum that is roughly
of the form 
\be
\omega= i n \left[{\rm real~number}\right]+\left[{\rm complex~number}\right]\;.
\label{I1}
\ee
Here,  $n=0,1,2,...$ is the discrete parameter labeling the 
modes,~\footnote{Note that  $n$, although discrete, 
is  {\it not} a quantum number {\it per se}.} 
and the unspecified numbers depend, in general, on both the
type of perturbation (primarily, on its spin and  angular momentum) and the  
black hole background.

Now specializing to  the case of a  highly damped scalar or gravitational
perturbation of a Schwarzschild black hole, one finds that, as
$n\rightarrow\infty$,~\footnote{The  relaxation
time of the perturbation goes as  the inverse of the imaginary 
part of the frequency. Hence, the  limit of high damping is synonymous
with the limit of large $n$. Further note that, although the highly damped
limit is of interest to the theoretical community, it is actually
the small $n$ modes which are most significant to gravitational wave
astronomers \cite{Kokkotas}.}
\be
\omega= i\kappa \left[n+{1\over 2}\right]+{\kappa\over 2\pi}\ln(3)
+{\cal O}[n^{-1/2}]
\;,
\label{I2}
\ee
where $\kappa$ is the (black hole) surface gravity.
Note that this result has been substantiated both numerically
\cite{Nollert-1,Andersson-1} and analytically \cite{Motl-2,Motl-1,Andersson-2},
and also applies to $d$-dimensional Schwarzschild black holes (with $d\geq 4$)
\cite{Motl-1,Birmingham-3,Cardoso}. 
Further note that the angular momentum of the perturbation
makes no contribution to the  finite-order terms. 

Another essential ingredient in   Hod's  conjecture is
Bekenstein's notion of a quantized black hole area spectrum 
\cite{Bekenstein-3}. To elaborate, Bekenstein has argued --- primarily,  
on the basis
of the horizon area being an adiabatic invariant \cite{Bekenstein-1} ---
that the  area ($A$) should have  a  quantum spectrum
of the form (in Planck units)
\be
A(m)=m \epsilon \quad {\rm where}\quad m=0,1,2,...\;,
\label{I3}
\ee
with $m$ designating the quantum number and
 $\epsilon$ being an undetermined constant of the order 
unity.~\footnote{There
has since been many studies that have argued for this 
spectral form using more rigorous methods. See \cite{Das} for a list
of references.} Assuming that  this is correct,
one might expect the precise  value of $\epsilon$ to be a ``residue'' of the
underlying fundamental theory. In particular, as
originally noted by Bekenstein and Mukhanov
\cite{Bekenstein-4}, the Bekenstein--Hawking entropy formula 
($S=A/4$ in Planck units) combined with a statistical interpretation of the
entropy 
requires that $e^{A/4}$ is an integer.
This, in turn, constrains the spacing to take the form $\epsilon=4\ln(k)$
for some integer $k$.

The Hod conjecture is based on the idea that the quasinormal
modes --- a purely classical effect --- can be used to fix, unambiguously,
the ``quantum-gravity  parameter''  $\epsilon$.  Hod's main point
is that, ``in the spirit of the Bohr correspondence principle'', 
the real part of the highly damped mode frequencies should
be interpreted as a characteristic transition frequency (say, $\omega_c$) 
for the black
hole \cite{Hod-1}; so that (for Schwarzschild anyways) 
$\omega_c=(\kappa/2\pi)\ln(3)$.  Given this identification, some
simple algebra then fixes  $\epsilon=4\ln(3)$.
The ``miracle'' is that this spacing, arrived at by purely
semi-classical arguments, agrees with the above 
 constraint (as motivated by quantum gravity)  {\it and} fixes
$k=3$. 

Remarkably, the above argument carries through,
unfettered, for higher-dimensional Schwarzschild black holes
(see \cite{Gabor-1} and \cite{Motl-2}). It is perhaps useful
for the ensuing discussion to show how Hod's conjecture can be
applied to generic Schwarzschild-like black holes. Let us
assume that there is only a single independent dimensionful
parameter needed to describe the black hole geometry, which 
can be taken without loss of generality to be the Hawking
temperature $T$. Then, on dimensional grounds, the real part of the
 large-damping
quasinormal-mode  frequency must be proportional to $T$:
\be
\omega_c={\epsilon\over4} T \;.
\ee
If $\omega_c$ is indeed a fundamental vibration frequency
associated with black holes, then standard arguments
(see \cite{Gabor-1} and references therein) imply that
the following is an adiabatic invariant with an equally
spaced spectrum at the semi-classical level:
\be
\int{dE\over \omega_c}= {4\over \epsilon}\int {dE\over T} 
= {4\over\epsilon}S \approx m \;,
\ee
where $E$ is the black hole energy and the thermodynamic first law has 
been applied.
Thus, the semi-classical spacing of the Bekenstein--Hawking entropy
(and, hence, the black hole area) 
is generically determined by $\omega_c$ to be
\be
S\approx {m\epsilon\over4} \;.
\ee
Given that $S$ has  a statistical-mechanical interpretation
in terms of black hole microstates, then  $S = m \ln(k)$,
which requires $\epsilon=4\ln(k)$. Note that this
argument requires only three things:
{\it (i)} Hod's conjecture that $\omega_c$ is a fundamental
oscillation frequency associated with the black hole, 
{\it (ii)} the first law of black hole thermodynamics,
and {\it (iii)} the existence of a only a single dimensionful parameter.
It is the absence of this last condition that makes it
difficult to apply the above argument directly to more
general types of black holes; such as black holes that 
are rotating  or charged, or those in a non-asymptotically flat 
spacetime.

We should note, in passing, that  the Hod conjecture has played a recent,
prominent role in attempts to fix the Immirzi parameter \cite{Immirzi} (which
is closely related to Bekenstein's $\epsilon$) in the area spectrum of
 loop quantum gravity (LQG) \cite{Ashtekar}. Most significantly, it has been 
suggested that the presence of $\ln(3)$ implies a change in the LQG
 gauge group
from $SU(2)$ to  $SO(3)$ \cite{Dreyer}. On the other hand, some very recent 
calculations of the Immirzi parameter \cite{Poland-1,Poland-2}
cast serious doubts on the feasibility of applying Hod's conjecture in 
this particular context.  As this affair remains currently  unsettled 
\cite{New-lqg}, we
will comment no further on the LQG connection at this time.

It is not our intent to debate, one way or the other, whether
the Hod conjecture truly represents a  duality between quasinormal
modes and quantum gravity or is, rather, just a numerical coincidence.
(For some discussion in support of the latter viewpoint, see 
\cite{Joey-1,Khriplovich}.) Nonetheless, one can reasonably argue that a 
minimal 
requirement for the former perspective is  some degree of universality;
inasmuch as  $\epsilon$ should not be a particularly model-specific
parameter
if truly representative of the fundamental theory.
Hence, one might hope that, at the very least, the
same value of $\epsilon$ would be  found for {\it any} single-horizon
black hole
in an asymptotically flat spacetime.~\footnote{It is now well known
that, for a black hole in an
 asymptotically  de Sitter or
anti-de Sitter spacetime, the quasinormal-mode spectrum  
 will be qualitatively much different than that of an 
asymptotically flat spacetime \cite{CNS}. It is up to the individual reader to
decide if the   Hod conjecture 
 should be required to persist  under such conditions.} 
If this were indeed the case, then ---  by virtue of the conjectured
duality ---  one would expect the quasinormal spectra of
 all such black holes  to 
limit to the same value of $\Re(\omega)/\kappa$.~\footnote{One
could legitimately argue that $\epsilon$ could still  be dependent on
the dimensionality ($d$)
without jeopardizing the conjectured  correspondence.
Such a dependence, however, does not occur for (at least) $d\geq 4$, 
since $\epsilon$ has already
been calculated to be $4\ln (3)$  for
$d$-dimensional Schwarzschild black holes \cite{Motl-1,Birmingham-3,Cardoso}.} 
Thus, in order to understand the physical significance (if any)
of this circle of ideas, it becomes important that the quasinormal modes
are investigated in as broad a context as possible.
Unfortunately, the various technical difficulties associated with 
quasinormal-mode calculations have impeded progress along such general 
lines.~\footnote{However, for some substantial progress at generality, see
\cite{Tamaki}, which extends the WKB analysis of \cite{Andersson-2}.}

Let us now turn our attention to 
two-dimensional theories of dilatonic gravity.  Two-dimensional
models have, often in the
past, served as
a useful framework for exploring
gravity-related issues; both at the semiclassical and quantum level. 
(For reviews and references, see \cite{Odintsov,Kummer}.) 
The aim of the current paper is to
use this lower-dimensional context as a means for calculating
the quasinormal-mode spectra for a wide class of single-horizon
black holes. (Dilatonic gravity can also be used, in principle, to study
more complicated types of black holes, but we will defer such endeavors until
a later time.) Let us point out that, although two-dimensional theories
of gravity  are interesting in  their own right, 
such models can also have
physical significance as representing reduced forms of 
higher-dimensional gravity.
For instance, a class of dilaton-gravity models 
arises from the spherically symmetric reduction of
$d$-dimensional Einstein gravity (with $d\geq 4$)
\cite{Gabor-2}, while another class arises from the axial
reduction \cite{Ortiz} of the ($d=3$) BTZ black hole \cite{BTZ}.

The rest of the paper proceeds as follows. In the next section, 
we introduce our generic dilaton-gravity model; discussing
the action, solutions and (briefly) the thermodynamics.
In Section III, we consider an appropriate wave equation
and cast it into a form that is suitable for the analysis of
the quasinormal modes.  The quasinormal spectrum is then calculated
(for a still quite general class of ``power-law'' theories) in   Section IV.   
More specifically, we are able to extend  the powerful 
 monodromy methodology of
\cite{Motl-1} into this dilaton-gravity framework.  
Section V contains  a summary and discussion of our results.
Finally, there is  an appendix, where  
we provide a more detailed account of the  monodromy calculation.

\section{The Model}

Let us begin by introducing our generic model of interest.
To be precise, we will consider the most general 1+1-dimensional action 
that depends
on a metric tensor and a dilaton scalar ($\Phi$), and complies 
with diffeomorphism invariance and no more than
two derivatives of the fields; that is \cite{Banks,Odin},
\be
I= {1\over 2G}
\int d^2x \sqrt{-{\overline g}}
\left[D(\Phi)R({\overline g}) + {1 \over 2}\left(\nabla \Phi\right)^2
+{V_{\Phi}(\Phi)\over l^2}\right] \;.
\label{1}
\ee
Here, $D$ and $V_{\Phi}$ are arbitrary  ({\it i.e.}, model-dependent)
functions of the dilaton, while $G$ (a dimensionless gravitational coupling) 
and  $l$ (some length scale) may be regarded as
fundamental parameters of the two-dimensional theory.

With the modest constraints that both $D(\Phi)$  and its derivative are 
non-vanishing functions, there exists a field reparametrization which
conveniently  eliminates
the kinetic term of the action \cite{Gabor-3}.  To elaborate,
by defining
\be
\Omega^2(\Phi) \equiv \exp\left[{1\over 2}\int 
{d\Phi\over \left(dD/d\Phi\right)}\right]\;,
\label{2}
\ee
\be
g_{\mu\nu}\equiv \Omega^2(\Phi){\overline g}_{\mu\nu}\;,
\label{3}
\ee
\be
\phi \equiv D(\Phi)\;,
\label{4}
\ee
\be
V_{\phi}\left[\phi(\Phi)\right] \equiv {V_{\Phi}(\Phi)\over \Omega^2(\Phi)}\;,
\label{5}
\ee
we then obtain 
\be
I= {1\over 2G}
\int d^2x \sqrt{-g}
\left[\phi R(g)
+{V_{\phi}(\phi)\over l^2}\right] \;.
\label{6}
\ee
Given that the model admits black hole solutions  --- which will always
be assumed here ---  it can be shown that any thermodynamic
property of the black hole is invariant under such a field 
reparametrization \cite{Gabor-4,Cadoni-1}.
Hence, it will be sufficient, for current considerations, to
work with  the revised form of the action (\ref{6}).

One of the virtues of the reparametrized action is that the
general solution can readily be obtained \cite{Gabor-4,Strobl}.  
More specifically, the gauge choice 
\be
\phi={x\over l}\geq 0 \;
\label{7}
\ee
leads to a (static) Schwarzschild-like metric,
\be
ds^2= - f(x)dt^2+ f^{-1}(x)
dx^2 \;,
\label{8}
\ee
where
\be
f(x)\equiv J(x)-2lGM\;.
\label{9}
\ee
Here,  $M\geq 0$ is a constant of integration that can be identified
with the  conserved mass  of the black hole \cite{Mann} and
\be
J[\phi(x)]\equiv \int^{\phi=x/l}V_{\phi}({\tilde\phi})d{\tilde\phi} \;,
\label{10}
\ee
with  the integration constant having already
been incorporated into  the observable $M$.

Although a fully  general treatment is (at least in principle) possible, 
we will often 
focus on a particular class of ``power-law potentials'';
namely,
\be
V_{\phi}\sim \phi^{-b}\;,
\label{11}
\ee
so that
\be 
J\sim \phi^{1-b}\;.
\label{12}
\ee
In concentrating on the power-law class, one of our motivations  is that it
 enables
a precise prescription for the admission of black hole solutions.
The relevant class of models turns out to be  
$-1\leq b < +1$ \cite{Cadoni-2}. 

Such power-law potentials can also be  motivated
from the perspective of ``physically relevant'' models;
by which we mean two-dimensional gravity models that are obtained 
from the dimensional reduction of a higher-dimensional theory.
Most notably, the spherically symmetric reduction of $n$+2-dimensional
Einstein gravity (with $n\geq2$) 
yields, after  the appropriate parametrizations, a power-law
potential with $b=1/n$ \cite{Gabor-2}. In fact, many (if not most)
of the models considered in the literature take on  a power-law
form [although not necessarily in compliance with equation (\ref{13}) below].
Other popular theories include Jackiw--Teitelboim gravity \cite{JT}
(or, equivalently, the dimensionally reduced BTZ black hole \cite{BTZ,Ortiz}) 
with $b=-1$, and the Weyl-rescaled CGHS model \cite{CGHS}
with $b=0$ \cite{Cadoni-3}.

For future reference, we will fix (without loss of generality)
the implied prefactor in equation (\ref{11}) equal
to $(1-b)$, so that $J=\phi^{1-b}$ and, consequently,
\be
f(x) = \left({x\over l}\right)^{1-b}-2lGM\;
\label{14}
\ee
describes the power-law metric. 
The black hole horizon   --- in general,
defined by the relation $J(x= x_h)=2GMl$ --- is now readily located:
\be
x_h = l [2lGM]^{1/1-b}\;.
\label{15}
\ee

As a related matter of interest, any  model that does admit black
hole solutions  has an associated 
temperature ($T$), surface gravity ($\kappa$)
and entropy ($S$), which are given by  \cite{Gabor-4}
\be
T={\kappa\over 2\pi}=\left.{1\over 4\pi}{df\over dx}\right|_{x=x_h}=
{1\over 4\pi l}V_{\phi}(\phi_h)\;,
\label{16}
\ee
\be
S={2\pi\over G}\phi_h\;,
\label{17}
\ee
where (of course) $\phi_h=x_h/l$. Note that the temperature follows from 
the usual Gibbons--Hawking prescription \cite{Hawking-3}, while
the entropy then follows from the  first law of 
thermodynamics.~\footnote{The entropy can also be calculated
with Wald's Noether charge formalism \cite{Wald}, which then --- via
the first law --- provides an independent verification for interpreting $M$
as the black hole mass.}
 It is also worth re-emphasizing that any of these 
thermodynamic properties is invariant under our prior field
reparametrization.

One final point concerning the thermodynamics:
Rewriting the entropy as $S\sim \phi_h \sim M^{1/(1-b)}$ [{\it cf}, equations
(\ref{7}) and (\ref{15})],  
we see that, if  the entropy is to grow with the mass, the
bound $b<1$ is required. Moreover, similarly re-expressing the
temperature as $T\sim \phi_h^{-b}\sim M^{-b/(1-b)}$,
we observe that, if the temperature (and surface gravity) 
is to decrease with  increasing energy,  the bound $b>0$
is required. (Although counter-intuitive, the resulting
negative heat
capacity is  the normal state of affairs for  a Schwarzschild
black hole \cite{Hawking-3}.) On this basis,
we will ultimately restrict considerations to the sub-class
of power-law potentials having
\be
0 < b < 1\;.
\label{13}
\ee
Furthermore,  given the power-law form, this upper
bound is, as previously mentioned,  necessary
for the existence of black hole solutions; whereas the lower bound
ensures that the physically relevant solutions of the ``fundamental'' action 
(\ref{1}) will be asymptotically flat.

\section{The Wave Equation}

Before embarking on a quasinormal-mode analysis of our dilaton-gravity
model, we first require a suitable
generalization of the (usual) Klein--Gordon equation.  Significantly, this
equation should  describe the dynamics of a matter perturbation in the 
two-dimensional 
black hole background [as described, in full generality, by equations
(\ref{7})-(\ref{10})].  
Let us allow  for an arbitrary
coupling [say, $h(\phi)$] between the dilaton and the
matter perturbation, then the  following can be viewed as a natural
generalization: 
\be
\nabla^2_{h}\Psi\equiv \partial_{\mu}\left[\sqrt{-g} h(\phi) g^{\mu\nu}
\partial_{\nu}\Psi\right]=0\;.
\label{18}
\ee
Here, the perturbation field, $\Psi=\Psi(x,t)$, is taken to be
a scalar (spinless) field.  This choice is motivated by
 two-dimensional gravity
being a topological field theory, and so there can be no
propagating physical modes; in particular,  no gravitons nor 
photons  \cite{Birmingham-2}.
(Note that, as is typical, we consider
the case of a massless perturbation.)

An essential feature of $\nabla^2_{h}$ in two dimensions is its
invariance under a conformal reparametrization of the metric.
This observation (along with the similar invariance of any thermodynamic 
property)
 tells us that our prior field redefinitions  will in
no way 
effect the physical spectrum of the quasinormal modes.

From a two-dimensional perspective, the coupling  $h(\phi)=h(x)$
could only be decided after one has explicitly specified the theory. 
On the other hand, 
when the model has a higher-dimensional pedigree,
the natural choice would be the part of  the (higher-dimensional)
metric determinant that is ``lost'' in the reduction 
procedure.~\footnote{This perspective follows from the presumption
of diffeomorphism invariance in the higher-dimensional theory.} 
For instance, in  spherically symmetric Einstein gravity,
one would be inclined to set $h$ equal to  $\sqrt{-g^{(n+2)}}=r^{n}$,
where $r$ is the (standard) Schwarzschild radial coordinate for
the $n$+2-dimensional theory.
As it happens, this choice translates into $h=\phi$ \cite{Gabor-2}.
Nonetheless, we will keep things rather general for the moment.

To obtain 
a operational form of wave equation, it is convenient
to {\it (i)} employ a separation-of-variables  technique
and {\it (ii)} introduce a generalized ``tortoise
coordinate''. The first step can be realized by expressing
the perturbation field as follows:
\be
\Psi(x,t)={\psi(x)\over \sqrt{h(x)}}e^{i\omega t}\;,
\label{19}
\ee
where $\omega$ is the frequency and the spatial wavefunction,
$\psi(x)$, has been defined in a way that ensures the simplest
possible wave equation.  In direct analogy
to Schwarzschild calculations, the second step is realized
by the relation
\be
dz={dx\over f(x)}\;,
\label{20}
\ee
where $z$ denotes the tortoise-like (spatial) coordinate
and $f(x)$ is the metric function of equation (\ref{9}).
Although $z$ can not be solved for in a closed form under generic 
circumstances, we can still make some pertinent observations:
$z$ must diverge  at the horizon (where $f\rightarrow 0$)
and, for a metric of the {\it asymptotic} form $f\sim x^{1-b}$,
$z$ will go as $x^{b}$ at spatial infinity.
This means that,  as long as 
$f\sim x^{1-b}$
with $b>0$ is  valid  as $x\rightarrow\infty$ 
[which can be viewed as a generalization of the lower bound in 
equation (\ref{13})~\footnote{Moreover, this   
constraint on  $f$  complies with our prior notion of asymptotic
flatness. See the final paragraph in Section II.}],
the  black hole exterior 
$x\in(x_h,\infty)$
maps into the entire real line $z\in(-\infty,+\infty)$.

Incorporating the above relations and  the static solution
(\ref{7})-(\ref{10}) into equation (\ref{18}),
we eventually obtain the following Schrodinger-like form (in direct analogy
to the Regge--Wheeler equation \cite{Kokkotas}):
\be
\partial^{2}_{z}\psi(z) -U[z(x)]\psi(z)=-\omega^2\psi(z)\;,
\label{21}
\ee
where the ``scattering potential'' is given by
\be
U(x) ={1\over 2}{f\over h}\left[f h^{\prime\prime}
+f^{\prime}h^{\prime}-{1\over 2}{f\over h}\left(h^{\prime}\right)^2\right]\;,
\label{22}
\ee
with a prime denoting a  differentiation with respect to $x$.

Significantly, we have attained
a one-dimensional scattering problem (mapped to the entire real line) 
with a ``mostly positive''
potential. (By ``mostly positive'' it is meant that the potential is positive
at least in the proximity of the horizon; this being the region of the
spacetime that is most relevant to the physical scattering problem 
\cite{Joey-3}.)
Note that it is the latter feature that leads to complex-frequency
solutions and, hence, the notion of {\it quasinormal} modes.

It is, of course,
the scattering potential, $U(x)$, that ultimately determines the spectral
features of these quasinormal modes.  For future purposes, 
let us specialize our
generic result (\ref{22}) to the power-law potential; that is,
we now set $V_{\phi}(\phi)=(1-b)\phi^{-b}$ or, equivalently,  
$f(x)=(x/l)^{1-b}-2lGM$ (with $0<b<1$).
As motivated below, we also set $h=\phi^a$  (with $a>0$)
to obtain
\be
U(x) ={af\over 4}\left[(a-2){f\over x^2}+{2\over l}{V_{\phi}\over x}\right]\;.
\label{23}
\ee
Regarding the matter-dilaton coupling,
it may not be {\it a priori} clear as to what form 
 $h(\phi)$  should adopt 
in this framework. Nevertheless, let us recall that, in the physically
motivated case of spherically reduced Einstein gravity (for
which $b=1/n$ in $n$+2 dimensions), the natural choice
appears to be $h=\phi$. On this basis, it would then seem reasonable
to use our power-law {\it ansatz} of  $h=\phi^a$.  To keep matters as general
as possible,
 we will  regard  $a$ as an 
arbitrary (coupling) parameter with the constraint that $a>0$; 
although keeping in mind that $a=1$
is a particularly well-motivated choice. (Note that $a=0$ coincides
with the trivial case of a vanishing potential. Further note that
$a<0$ will 
jeopardize the positivity of the potential in the critical near-horizon 
region.~\footnote{Actually, 
this positivity may not be immediately
apparent for $0<a<2$. Nonetheless, the sceptical reader should notice that,
for this range of $a$ values, the second (intrinsically
positive) term in equation (\ref{23}) will inevitably win out near
the horizon (where $f\rightarrow 0$).})

\section{The Quasinormal Modes}

Taking  equations (\ref{21}) and (\ref{23}) as our starting point, 
we will now  investigate the asymptotic nature of
the associated quasinormal-mode spectrum. But, before getting into specifics,
some general commentary on the nature of the problem (including the boundary 
conditions) 
is in order. 

Because of our physically motivated restrictions on the solution
({\it i.e.}, $0<b<1$ and $a>0$), it is clear from equation (\ref{23})
that the scattering potential has the following properties:
{\it (i)} it is mostly  positive (in the sense described above)
and {\it (ii)} it decays to zero at both the horizon and asymptotic infinity.
[To be explicit, the asymptotic form as $x\rightarrow\infty$ is $U\approx
a(a-2b)/(4l^2 b^2 z^2)$ with $z\sim x^b$.]
On the basis of these properties --- reminiscent of a highly damped
system --- one anticipates that there
will be no normalizable bound states. However, there can still be 
  a set of discrete states with complex-valued frequencies; these being the
so-called quasinormal modes. Essentially, each of these modes is identifiable
with a pole in the scattering amplitude (see \cite{Kokkotas} for further 
background).

Since  $U[z(x)]$ goes to zero at both the horizon  
($z\rightarrow -\infty$) and spatial infinity ($z\rightarrow\infty$),
it immediately follows from equation (\ref{21}) that
\be
\psi(z)\sim e^{\pm i\omega z}\quad {\rm as} \quad z\rightarrow\pm\infty\;.
\label{24}
\ee
One normally proceeds by imposing  ``radiation'' boundary conditions,
so that the mode is a purely ingoing plane wave at the horizon and
purely outgoing  at infinity. Considering our prior convention
of $\Psi\sim e^{+i\omega t}$, the appropriate boundary conditions
become
\be
\psi(z)\sim e^{+i\omega z}\quad {\rm as} \quad z\rightarrow -\infty\;,
\label{25}
\ee
\be
\psi(z)\sim e^{-i\omega z}\quad {\rm as} \quad z\rightarrow +\infty\;.
\label{26}
\ee
Note that  any damped mode 
must decay with time [and so  $\Im(\omega) > 0$]; meaning that
the ``absent components'' (outgoing at the horizon and ingoing at
infinity) would have to  be exponentially suppressed anyways. This subtlety
(amongst others) has impeded general progress in 
quasinormal mode calculations.
One possible resolution, as proposed in \cite{Motl-1}, is to
analytically continue $z$, instead of $\omega$, and then consider the 
boundary conditions as the {\it product} $\omega z\rightarrow\pm\infty$. 
Consult \cite{Motl-1} for further discussion. 

There are, by now, many documented techniques for calculating the 
quasinormal-mode frequencies; both numerical and analytical. (For an up-to-date
survey, see \cite{Cardoso}.) Here, we will focus on an analytical
  method that is particularly well suited
for determining the frequencies in the highly damped
[$\Im(\omega)\rightarrow\infty$] limit; namely,  the  
 ``monodromy procedure'' of Motl and Neitzke \cite{Motl-1}.
In a nutshell, their methodology  is based (in a Schwarzschild framework) on 
considering the complex-$r$ plane (including the black hole
interior) and then 
 computing the monodromy
that the wavefunction picks up (relevant to the singular horizon point)
in two distinct ways.  These two calculations can then be matched --- 
in the  limit of high damping  --- resulting in
an equation from which  the quasinormal-mode spectrum can 
readily be  extracted.
It is encouraging that, for the four-dimensional Schwarzschild 
black hole, their outcome agrees
with earlier numerical calculations \cite{Nollert-1,Andersson-1}.

Although we  intend to adapt this monodromy method to 
our generic dilaton model, it will (thankfully) not be necessary to
reiterate the rather technical procedure. 
(We do, however, provide a somewhat 
technical outline of the procedure in an appendix.
The interested reader can also consult
the original work \cite{Motl-1} for further details.)  
Instead, we will demonstrate the similarities between our problem
and the Schwarzschild one, and use this as a basis for suitably
generalizing the results of \cite{Motl-1}.

First of all, let us point out that the complex-$r$ plane of
 the Schwarzschild black hole is qualitatively  similar
to the complex-$x$ plane of our dilatonic black hole model.
In particular, the key elements in common are two regular singular points
at the spatial origin and the black hole horizon and an irregular singular
point at spatial infinity. Note that the calculation is insensitive 
to any subtle differences that may exist.  Actually, what is
important to the monodromy calculation 
is the near-origin form of the scattering potential 
when expressed in terms of the 
tortoise coordinate; this point is elaborated on
 in section 3.1 
of \cite{Motl-1} (where the  authors generalize considerations to 
higher-dimensional Schwarzschild black holes --- also see 
\cite{Birmingham-3}). 

In view of the above observation, let us see what does 
happen to be the behavior of our scattering potential
near the spatial origin; especially, its functional dependence
on the generalized tortoise coordinate [{\it i.e.},  
 $z$  as defined by equation (\ref{20})].
To proceed, it will be  necessary for us to take the near-origin limit
of a few quantities.  First consider that (near $x = 0$) 
$f\approx -2lGM$. Then, since $V/x\sim x^{-(b+1)}$ ``blows up''  slower
than $x^{-2}$,  equation (\ref{23}) gives us
 \be
U(x\sim 0)\approx {a(a-2)(2lGM)^2\over4x^2}\;.
\label{27}
\ee

Now consider that, in the near-origin limit, equation (\ref{20})
can be trivially integrated to yield $x=-(2lGM) z$.~\footnote{As in
\cite{Motl-1}, the
constant of integration is   chosen 
so that $z=0$ at the origin.}  Hence,
the above expression for $U$ can be rewritten as 
\be
U(z\sim 0)\approx {a(a-2)\over 4z^2}\;.
\label{28}
 \ee
Remarkably, this is identical to the near-origin form
of the potential derived in \cite{Motl-1} [{\it cf}, their
equation (13)], insofar as we make the identification 
\be 
a=1\pm j\;,
\label{29}
\ee
where $j$ is the spin associated with the Schwarzschild perturbation.
(Note that, typically,  $j=0,1,2$, {\it etc.} --- so that,
given $j$, we can choose the sign 
by  requiring that $a>0$.)

In view of these observations, the monodromy calculation
of \cite{Motl-1} should carry through, almost {\it verbatim},
 provided that we make the specified
substitution. In this spirit, we have obtained (see the {\it appendix}  for
an elaboration) the following
relation:
\be
e^{2\pi\omega /\kappa}=-\left(1 + 2\cos[\pi(a-1)]\right)\;, 
\label{30}
\ee
where $\kappa$ is the surface gravity of the dilatonic black hole [{\it cf}, 
equation (\ref{16})]. It is worth emphasizing
that  $\kappa$ appears because of  a
contour integral around the event horizon. Moreover, the appearance of
the surface gravity  is, in fact,
a {\it universal} consequence of the {\it near-horizon} form of the 
tortoise coordinate.  By which we mean that
\be
z=\int{dx\over f(x)}\approx\int{dx\over(x-x_h)f^{\prime}(x_h)}
={1\over 2\kappa}\ln(x-x_h)
\label{31}
\ee
is universally valid. 
(Also see equation (47) of \cite{Motl-1} along with the related discussion.)

Another interesting feature of the relation (\ref{30}) is that the right-hand
side depends 
{\it only}
on the value of $a$; namely, a parameter that determines
the coupling between the dilaton and the matter perturbation.
Which is to say, the quasinormal modes seem rather indifferent
to the form of the gravitational action itself; hence, we can see
the first glimpses of the universality
that is suggested by the Hod conjecture \cite{Hod-1}.

Although $a$ was left arbitrary in our formalism, two results
immediately jump out. Firstly, when $a$ is 
any even integer, the right-hand side
of this expression becomes unity, and it follows that the real
part of each  mode frequency vanishes asymptotically.  This is similar
to what is found for electromagnetic ($j=1$) perturbations 
in many black hole models (including Schwarzschild).
Meanwhile, when $a$ is any odd integer, the right-hand side becomes
$-3$. Solving this relation, we find that
\be
\omega =   i\kappa\left[n+{1\over 2}\right]+ {\kappa\over 2\pi}\ln (3)\;,
\label{32}
\ee
where $n$ takes on positive integral values (the positivity is
enforced by the boundary condition that $\Im(\omega) >0$).  Actually, this
 is not
quite right because, as previously mentioned, the monodromy calculation
only has validity  in the asymptotic limit of high damping. Hence, it
is more accurate to write
\be
\omega =   i\kappa\left[n+{1\over 2}\right]+ {\kappa\over 2\pi}\ln (3)
+{\cal O}[n^{-1/2}]\quad {\rm as}\quad n\rightarrow\infty
\;.
\label{33}
\ee
But, any way you slice it, this is precisely the highly damped
spectrum that is obtained for scalar ($j=0$) or gravitational ($j=2$)
perturbations of  a $d$-dimensional
 Schwarzschild black hole (with $d\geq4$).    In particular, $a=1$
--- that is, the preferred choice  of coupling parameter
as motivated by spherically symmetric  Einstein gravity ---   
gives us precisely the Schwarzschild spectrum. This
``coincidence'' provides a nice consistency check for
our calculation.

\section{Discussion}

In summary, we have considered a generic model of two-dimensional
dilatonic  gravity coupled to a (scalar) matter perturbation.
Most importantly, we have
 utilized the monodromy technique  of 
\cite{Motl-1} to calculate the frequency of the highly damped
quasinormal modes; including the (sub-leading) real part.
For the particular case of spherically symmetric reduced Einstein gravity,
we were able  to substantiate that the real part
goes asymptotically to $T\ln(3)$; in compliance with previous calculations
for a $d$-dimensional  Schwarzschild black hole 
\cite{Motl-1,Birmingham-3,Cardoso}.

An interesting subplot is the direct connection between $a$
(a parameter which determines the two-dimensional  matter-dilaton coupling)
and $j$ (the spin of the higher-dimensional matter perturbation);
{\it cf}, equation (\ref{29}).  Consequently, we can effectively
mimic different types of perturbations by judiciously selecting
the  parameter  $a$; even though, typically speaking,
 two-dimensional gravity 
can only support scalar ($j=0$)  perturbations. It might
prove to be informative if this connection could be given a physical
motivation.

Although the agreement with known results is quite gratifying, our
formalism is making a much more powerful statement than just this. 
To elaborate,
one might expect that virtually any single-horizon and
asymptotically flat black hole spacetime would have a dimensionally reduced
form  that complies  (after a suitable 
process of reduction and reparametrization)
with the model we have been studying. Then, since the final result for
the mode spectra depends only on $\kappa$ and $a$, it is feasible that, 
for any such black hole,
the  real part of the frequencies may asymptote to $T\ln(3)$. If this were 
indeed the case,
it would establish  conjectures of universality that have been made in, 
for instance, 
\cite{Gabor-1,Tamaki}. This type of  universality should  be regarded as
 a minimal
requirement if one is to take the Hod conjecture \cite{Hod-1} seriously
as a statement about quantum gravity. 

On the other hand (staying with the Hod conjecture), 
our calculation 
enables  one to  see, quite clearly,
 where the relevant contributions to the mode spectra 
come from. For instance, the exponent in equation (\ref{30}) or 
$2\pi \omega \kappa$ ({\it i.e.}, the surface gravity) 
 comes directly from a contour integral
around the event horizon.  Moreover,
the $\ln(3)$ comes from the matter-dilaton coupling, with the choice
of $a=1$ (for spherically reduced Einstein gravity) being a direct
consequence of the diffeomorphism invariance of the higher-dimensional theory.
With this knowledge in hand,  there no longer seems to be anything 
``mystical'' 
about the presence of  $\ln(3)$. In fact, it becomes 
difficult to see where there
could be any deep connection with quantum gravity!

Irrespective of  such speculations about quantum gravity, 
it would be interesting
to see how much more can be deduced about quasinormal
modes in the lower-dimensional framework of  dilatonic gravity.
The power of two-dimensional gravity is that it allows one
to examine a large class of theories with a single  and relatively simple 
model;
with the underlying diversity being reflected in  a few adjustable parameters 
(such as our power-law coefficient $b$).
For starters, we hope to extend the analysis beyond this simple power-law class
of potentials; part of our motivation for maintaining generality throughout
much of the analysis. In this regard, it should, however, be pointed
out that our quasinormal-mode analysis depends (for the most part) only on 
the behavior of the tortoise coordinate at the horizon, the origin and
 spatial infinity.
Which is to say, one can readily confirm that the prior analysis carries
 right through
 as long as, for a given model, the following conditions are met:  
{\it (i)} $V_{\phi}$ maintains the necessary asymptotic behaviors;
namely, $V_{\phi}\sim \phi^{-b}$ with $b<1$ near the origin 
[thus ensuring $U(\sim 0)\sim z^{-2}$]
 and  $V_{\phi}\sim \phi^{-b}$ with
 $b>0$ at spatial infinity [thus ensuring $z\rightarrow \infty$], 
and {\it (ii)} $V_{\phi}$
 is a monotonic function of $\phi$ (because, given the correct asymptotic
behaviors,
this would be sufficient to ensure a single-horizon black hole
solution).

However, it is not so trivial to generalize matters 
to  
dilaton-gravity models with two conserved charges (or, equivalently,
spacetimes with two horizons).
Such models arise quite naturally when one considers, for instance, spherically
reduced Einstein--Maxwell gravity \cite{Gabor-5,Joey-2}.
Progress along these lines could certainly help us to better understand
the physical meaning of the Hod conjecture and, perhaps, quantum
gravity in general.

Finally, let us end with  a discussion on what is a {\it generic} issue  of
the monodromy procedure.
At an intermediate step in the calculation, one finds
that the formalism  breaks down for a scalar
perturbation ($j=0$) in higher-dimensional studies \cite{Motl-1} or
for an $a=1$ coupling in ours. [See the note following
equation (\ref{AA5}).]
This is, however,  not
really a problem, as long as one understands that
the  $j=0$ or $a=1$ case  is to
be handled in  a ``limiting sense'' 
\cite{Krasnov}.
  Nevertheless, one might well
ask if there is some sort of physical significance to
this type of occurrence.
 
To address this last point, let us make a pertinent observation
as follows: If we redefine the spatial wavefunction so that
\be
\chi(x)\equiv \sqrt{f(x)}\psi(x) \;,
\label{A1}
\ee   
then the scattering equation (\ref{21}) can be recast into the 
form~\footnote{Note that this form can also be used as the basis for a
WKB analysis of the quasinormal modes \cite{Andersson-2} 
(also see \cite{Tamaki}). The full WKB analysis for 
the case of generic two-dimensional dilatonic  gravity will be presented 
elsewhere.}
\be
\chi^{\prime\prime} + R(x)\chi=0\;,
\label{A2}
\ee
where 
\be
R(x)={1\over f^2}\left[\omega^2-U(x)+{\left(f^\prime\right)^2\over
 4}-{f f^{\prime\prime} \over 2}\right]\;.
\label{A3}
\ee
Now specializing to our power-law model, we find that, near the origin,
\be
R(x\sim 0)\approx -{a(a-2)\over 4x^2}\;.
\label{A7}
\ee
Hence, the near-origin form of the (revised) scattering equation
(\ref{A2}) can be written as
\be
H\chi(x)=0\;,
\label{A8}
\ee
with the ``Hamiltonian'' operator  being expressible as
\be
H=-{1\over 2}{d\over dx^2}+ {g_c\over 2 x^2}\;,
\label{A9}
\ee
where we have defined a  ``coupling constant'',  $g_c\equiv a(a-2)/4$. 
Remarkably, this is exactly the form of
the Hamiltonian operator for conformal quantum mechanics
\cite{Conformal}.~\footnote{It should be noted that the 
same observation about conformal quantum mechanics also follows
from 
the $z$-coordinate form of the scattering equation;
{\it cf}, equation (\ref{28}). Nonetheless, 
we have chosen to work in terms of the coordinate $x$  because 
the above formalism makes it quite clear as to
how the Hamiltonian can take on the  conformal form  
both near the origin {\it and}
at the {\it horizon};  {\it cf}, the third term on the right-hand
side of equation (\ref{A3}).} 
  Moreover, the well-known critical value 
of the coupling constant --- $g_c=-{1/4}$ ---  occurs when
$a=1$; precisely at the same point where the limiting
procedure is required.

We expect that the appearance of conformal quantum mechanics
is more than a mere coincidence.  Indeed, this  same form of 
Hamiltonian appears when a scalar probe is used to investigate 
the near-horizon dynamics of a Schwarzschild black hole 
\cite{Vaidya,Birmingham}; clearly, a dynamical framework that is 
closely related 
to the quasinormal modes. Furthermore,
 as speculated in \cite{Birmingham},
there is probably a deep connection between this realization of conformal
mechanics and
calculations of the black hole entropy that are based on
near-horizon conformal symmetries \cite{Carlip,Solodukhin}.
We hope to readdress such fascinating ideas at a future time.

\section*{Appendix: The Monodromy Calculation}

Here, we provide a more detailed account of how 
equation (\ref{30}) is obtained by way of the monodromy approach.
We will, however, provide only a somewhat brief sketch
of the computation; glossing over many of the technical
caveats. The reader is referred, once again, to \cite{Motl-1}
for many of the finer subtleties of the calculational procedure.

The  basic idea is to impose  the highly damped limit or
 $\Im(\omega) >> \Re(\omega)$ (with both quantities taken to be positive) and 
then
 calculate the monodromy of the wavefunction
in two distinct ways. These  two results can  then be matched  to obtain 
the desired relation.  

We begin here by considering a suitably constructed contour in the 
complex-$x$ plane
or,  more accurately, the complex-$z$ plane.  However, before getting started,
we will (for ``best results'')
perform a  ``Wick rotation'' of  $z$,
so that  the boundary condition at spatial infinity
($z$ or $x\rightarrow\infty$) now
becomes $\omega z \rightarrow\infty$.
 It is  convenient if the contour is initiated in this  
particular asymptotic region --- effectively, this means  starting
on the negative imaginary axis of $z$ --- where we
 know [{\it cf}, equation (\ref{26})] that 
\be
\psi(z)\sim e^{-i\omega z} \quad{\rm as} \quad \omega z\rightarrow +\infty\;.
\label{AA1}
\ee

Next, we  follow the line
$\Im(\omega z) =0$  until the near-origin region is reached.
 (Note that the plane-wave behavior should prevail until this interior 
region, since $\omega^2$ will dominate over the scattering 
potential up to a small neighborhood
of the origin.)  Let us appropriately recall the near-origin form
of the scattering equation [{\it cf}, equations (\ref{21}) and (\ref{28})], 
\be
\left[-\partial^{2}_{z} + {a(a-2)\over 4z^2} -\omega^2\right]\psi(z)= 0
\quad {\rm as} \quad z\rightarrow 0\;.
\label{AA2}
\ee
Conveniently,  this equation can be solved exactly as a linear combination
of  Bessel functions \cite{Bessel}:  
\be
\psi(z)=A_{+}c_{+}\sqrt{\omega z}J_{+\nu}(\omega z)
+A_{-}c_{-}\sqrt{\omega z}J_{-\nu}(\omega z)
\;, 
\label{AA3}
\ee
where  $\nu \equiv (a-1)/2$,  while the products  $c_{+}A_{+}$ and $c_{-}A_{-}$
represent constant coefficients.
Following \cite{Motl-1}, we
will choose the ``normalization factors'' (denoted by $c_{\pm}$) so
that
\be
c_{\pm}\sqrt{\omega z}J_{\pm\nu}(\omega z)
\sim 2\cos\left(\omega z-\alpha_{\pm}\right)
\quad {\rm as} \quad\omega z\rightarrow\infty\;,
\label{AA4}
\ee
with
\be
\alpha_{\pm}\equiv{\pi\over 4}\left[1\pm 2\nu\right]
={\pi\over 4}\left[1\pm(a-1)\right] \;.
\label{AA5}
\ee
Note that, when $\nu=0$, 
the Bessel function  is simply a constant and
equation (\ref{AA4}) is, strictly speaking, no longer accurate. 
Hence, when applied to the case of $a=1$, the subsequent formalism should be
 understood in a ``limiting sense'' \cite{Krasnov}.

Using equations (\ref{AA3}) and (\ref{AA4}), as well as the boundary
condition of equation (\ref{26}) (which indicates that the coefficient
of $e^{+i\omega z}$ must vanish as $\omega z\rightarrow \infty$),
we can deduce the constraint
\be
A_+e^{-i\alpha_{+}}+A_-e^{-i\alpha_{-}}=0\;.
\label{AA6}
\ee
Reapplying equations (\ref{AA3}) and (\ref{AA4}),  we also find that
\be
\psi(z)\sim \left[A_+e^{+i\alpha_{+}}+A_-e^{+i\alpha_{-}}\right]e^{-i\omega z}
\quad {\rm as}\quad \omega z\rightarrow \infty \;.
\label{AA7}
\ee

Let us now continue  the contour in the  following manner:  We want
to extend the contour from 
the negative imaginary axis of $z$ ($\Im (\omega z)\rightarrow\infty$)
to the positive imaginary axis ($\Im (\omega z)\rightarrow -\infty$)
while missing the origin.
As it turns out,
this action necessitates  a  counter-clockwise rotation of $3\pi$
about the singularity \cite{Motl-1,Birmingham-3,Krasnov}.  
[In determining this angle,
we are appealing to the structural similarity  between our model
and the Schwarzschild one; especially with
regard to the complex-$z$ plane in the limit of high damping.
More to the point, when $\left|\omega\right|^2$
is large enough  --- thereby dominating   the scattering problem except
in a small neighborhood of the origin ---  then only the near-origin
form of the potential is important. Relevantly, the near-origin
form of our potential is identical to that of the Schwarzschild model
when each is expressed in terms of their respective tortoise coordinates. 
This means that, essentially, any (hypothetical) observer ``living''
 in the complex
$z$-plane would be unable to distinguish between  the highly damped
limits of either scattering problem.~\footnote{Note 
that, by dealing directly with
the complex-$z$ plane, we circumvent any subtle differences that
might exist between our complex-$x$ plane and
the Schwarzschild complex-$r$ plane.}] 
Using some well-documented transformation properties of
the Bessel functions \cite{Bessel} to handle the rotation, we 
eventually arrive at 
\be
c_{\pm}\sqrt{\omega z}J_{\pm\nu}\sim e^{6i\alpha_{\pm}}
2\cos\left(-\omega z-\alpha_{\pm}\right)\quad {\rm as}\quad
\omega z\rightarrow -\infty
\;.
\label{AA8}
\ee
By way of equations (\ref{AA3}) and (\ref{AA8}), we then have
\be
\psi(z)\sim \left[A_+e^{5i\alpha_{+}}+A_-e^{5i\alpha_{-}}\right]e^{-i\omega z}
+\left[A_+e^{7i\alpha_{+}}+A_-e^{7i\alpha_{-}}\right]e^{+i\omega z}
\quad {\rm as} \quad \omega z\rightarrow -\infty
\;.
\label{AA9}
\ee

Finally, the contour is continued (clockwise)
 along a large semicircle --- encircling
the horizon ---
that  brings us from the asymptotic region  $\omega z\rightarrow -\infty$
back  to our starting point of $\omega z\rightarrow \infty$.
Significantly, the semicircle is strictly in the region 
where $\omega^2$ dominates over  the potential, and so
plane-wave behavior again prevails.  Given our limit of infinitely 
large damping (meaning that a WKB type
of approximation for the scattering problem must certainly be valid),
it follows that,
after completing this contour, the coefficient of the asymptotically dominant
 plane wave  ($e^{-i\omega z}$) should remain unchanged.~\footnote{Conversely,
 the same claim can not be made about the coefficient
of  $e^{+i\omega z}$, which is exponentially suppressed and, therefore,
much more sensitive to small corrections in a  WKB-like approximation.}
Hence, this outgoing plane wave picks up the following monodromy
about the specified contour [{\it cf}, equations (\ref{AA7}) and (\ref{AA9})]:
\be
{A_+e^{5i\alpha_{+}}+A_-e^{5i\alpha_{-}}\over
A_+e^{i\alpha_{+}}+A_-e^{i\alpha_{-}}} \;.
\label{AA10}
\ee
Using the boundary constraint (\ref{AA6}) to
eliminate the constants $A_{\pm}$ and then simplifying, one
finds that equation ({\ref{AA10}) reduces to
\be
{e^{6i\alpha_{+}} - e^{6i\alpha_{-}}\over
e^{2i\alpha_{+}} - e^{2i\alpha_{-}}} = -{\sin(3\pi\nu)\over
\sin (\pi\nu)}= -\left[1+2 \cos(2\pi\nu)\right]\;.
\label{AA11}
\ee

The above analysis  completes one way of calculating the monodromy. 
A second way
follows from the observation 
that the only singularity of $\psi(z)$ inside our chosen contour
occurs at the horizon.  So let us consider  the near-horizon form
of the wavefunction [{\it cf}, equations (\ref{25}) and (\ref{31})],
\be
\psi(z)\sim e^{+i\omega z}\sim \exp\left[i\omega\ln(x-x_{h})/(2\kappa)\right]
\quad {\rm as} \quad z\rightarrow -\infty\;.
\label{AA12}
\ee
This form tells us that going once around the horizon singularity
clockwise ({\it i.e.}, our chosen contour), multiplies
the wave function by $e^{2\pi\omega/(2\kappa)}$. However, the same
action multiplies the outgoing wave ($e^{-i\omega z}$)  by
$e^{-2\pi\omega/(2\kappa)}$. That is to say, the coefficient 
of $e^{-i\omega z}$ must be multiplied by a monodromy of
\be
{e^{2\pi\omega/(2\kappa)}\over e^{-2\pi\omega/(2\kappa)}}
=e^{2\pi\omega/\kappa}\;.
\label{AA13}
\ee

Finally, matching the two calculations for the monodromy (\ref{AA11})
and (\ref{AA13}) and then inserting $\nu=(a-1)/2$, we obtain
our spectral-defining relation; namely,  equation (\ref{30}).

\section*{Acknowledgments} 
\par

 Research for AJMM is supported by
the Marsden Fund (c/o the New Zealand Royal Society) 
and by the University Research Fund (c/o Victoria University).
Research for GK is supported by the Natural Sciences and Engineering Research
Council of Canada.
\par


\vspace*{20pt}

\begin{thebibliography} {99}


\bibitem{Hawking-1} S.W. Hawking, Phys. Rev. {\bf D14}, 2460 (1976).
\bibitem{Bekenstein-1} J.D. Bekenstein, Lett. Nuovo. Cim. {\bf 4}, 737 (1972);
Phys. Rev. {\bf D7}, 2333 (1973); Phys. Rev. {\bf D9}, 3292 (1974).
\bibitem{Hawking-2} S.W. Hawking, Nature {\bf 248}, 30 (1974);
 Comm. Math. Phys. {\bf 43}, 199 (1975).
\bibitem{Giddings} S.B. Giddings and N. Lippert, Phys. Rev. {\bf D69},
124019 (2004) [hep-th/0402073].
\bibitem{Fursaev} D.V. Fursaev, ``Can one understand black hole entropy without
knowing much about quantum gravity?'', arXiv:gr-qc/0404038 (2004).
\bibitem{Helfer} A.D. Helfer, Rept. Prog. Phys. {\bf 66}, 943 (2003)
[gr-qc/0304042].
\bibitem{Bekenstein-2} J.D. Bekenstein, Phys. Rev. {\bf D23}, 287 (1981).
\bibitem{tHooft} G. 't Hooft, ``Dimensional Reduction in Quantum
Gravity'', arXiv:gr-qc/9310026 (1993).
\bibitem{Susskind} L. Susskind, J. Math. Phys. {\bf 36}, 6377 (1995)
[hep-th/9409089].
\bibitem{Hod-1} S. Hod, Phys. Rev. Lett. {\bf 81}, 4293 (1998)
[gr-qc/9812002].
\bibitem{Kokkotas} K.D. Kokkotas and B.G. Schmidt, Living Rev. Rel.
{\bf 2}, 2 (1999) [gr-qc/9909058].
\bibitem{Bachelot} A. Bachelot and A. Motet-Bachelot, ``Resonances of
Schwarzschild black holes'', in {\it Proceedings of the IV International
Conference of Hyperbolic Problems}, ed. Vieweg (Taosmina, 1992).
\bibitem{Nollert-1} H.-P. Nollert, Phys. Rev. {\bf D47}, 5253 (1993).
\bibitem{Andersson-1} N. Andersson, Class. Quant. Grav. {\bf 10}, L61 (1993).
\bibitem{Motl-2} L. Motl, Adv. Theor. Math. Phys. {\bf 6}, 1135 (2003)
[gr-qc/0212096].
\bibitem{Motl-1} L. Motl and A. Neitzke, Adv. Theor. Math. Phys. {\bf 7},
307 (2003) [hep-th/0301173].
\bibitem{Andersson-2} N. Andersson and C.J. Howls, Class. Quant. Grav.
{\bf 21}, 1623 (2004) [gr-qc/0307020].
\bibitem{Birmingham-3} D. Birmingham, Phys. Lett. {\bf B569}, 199
(2003) [hep-th/0306004].
\bibitem{Cardoso} V. Cardoso, J.P.S. Lemos and S. Yoshida, Phys. 
Rev. {\bf D69}, 044004 [gr-qc/0309112].
\bibitem{Bekenstein-3} J.D. Bekenstein, Lett. Nuovo Cimento {\bf 11}, 467
(1974).
\bibitem{Das} S. Das, H. Mukhopadhyay and P. Ramadevi, ``Spectrum
of rotating black holes and its implications for Hawking radiation'',
arXiv:hep-th/0407151.
\bibitem{Bekenstein-4} J.D. Bekenstein and V.F. Mukhanov, Phys. Lett. 
{\bf B360}, 7 (1995) [gr-qc/9505012].
\bibitem{Gabor-1} G. Kunstatter, Phys. Rev. Lett. {\bf 90}, 161301 (2003)
[gr-qc/0212014].
\bibitem{Immirzi} G. Immirzi, Nucl. Phys. Proc. Suppl. {\bf 57}, 65 (1997)
[gr-qc/9710014].
\bibitem{Ashtekar} A. Ashtekar, J.C. Baez and K. Krasnov, Adv. Theo. Math.
Phys. {\bf 4},1 (2000) [gr-qc/0005126].
\bibitem{Dreyer} O. Dreyer, Phys. Rev. Lett. {\bf 90}, 081301 (2003)
[gr-qc/0211076].
\bibitem{Poland-1} M. Domagala and J. Lewandowski, ``Black hole
entropy from quantum geometry'', arXiv:gr-qc/0407051 (2004).
\bibitem{Poland-2} K.A. Meissner, ``Black hole entropy in loop
quantum gravity'', arXiv:gr-qc/0407052 (2004).
\bibitem{New-lqg} S. Alexandrov, ``On the counting of black hole
states in loop quantum gravity'', arXiv:gr-qc/0408033 (2004).
\bibitem{Joey-1} A.J.M. Medved and D. Martin, ``A note on quasinormal
modes: A tale of two treatments'', arXiv:gr-qc/0311086 (2003) and
 to appear in the conference proceedings for ACGR4.
\bibitem{Khriplovich} I.B. Khriplovich, ``Quasinormal modes, quantized
black holes, and correspondence principle'', arXiv:gr-qc/0407111 (2004).
\bibitem{CNS} V. Cardoso, J. Natario and R. Schiappa, 
``Asymptotic quasinormal frequencies for black holes in non-asymptotically
flat spacetimes'', arXiv:hep-th/0403132 (2004).
\bibitem{Tamaki} T. Tamaki and H. Nomura, ``The universal area spectrum
in single-horizon black holes'', arXiv:hep-th/0405191 (2004).
\bibitem{Odintsov}  S. Nojiri and S.D.
Odintsov, Int. J. Mod. Phys. {\bf A16}, 1015 (2001) [hep-th/0009202].
\bibitem{Kummer}  D. Grumiller, W. Kummer 
and D.V. Vassilevich, Phys. Rept. {\bf 369}, 327 (2002) [hep-th/0204253].
\bibitem{Gabor-2} G. Kunstatter, R. Petryk and S. Shelemy,
Phys. Rev. {\bf D57}, 3537 (1998) [gr-qc/9709043]. 
\bibitem{Ortiz} A. Achucarro and M.E. Ortiz, Phys. Rev. {\bf D48}, 3600
(1993) [hep-th/9304068].
\bibitem{BTZ} M. Banados, 
C. Teitelboim and J. Zanelli, Phys. Rev. Lett.
{\bf 69}, 1849 (1992) [hep-th/9204099].
\bibitem{Banks} T. Banks and M. O'Loughlin, Nucl. Phys. {\bf B362}, 649 (1991).
\bibitem{Odin} S.D. Odintsov and I. Shapiro, Phys. Lett. {\bf B263}, 183
(1991).
\bibitem{Gabor-3}  D. Louis-Martinez, J. Gegenberg and G. Kunstatter,
Phys. Lett. {\bf B321}, 193 (1994) [gr-qc/9309018].
\bibitem{Gabor-4} J. Gegenberg, G. Kunstatter and D. Louis-Martinez,
Phys. Rev. {\bf D51}, 1781 (1995) [gr-qc/9408015].
\bibitem{Cadoni-1} M. Cadoni, Phys. Lett. {\bf B395}, 10 (1997)
[hep-th/9610201].
\bibitem{Strobl} T. Strobl,  {\it Poisson Structure Induced Field
Theories and Models of 1+1 Dimensional Gravity},
PhD Thesis at TU-Vienna (1994) [hep-th/0011248].
\bibitem{Mann} R.B. Mann, Phys. Rev. {\bf D47}, 4438 (1993)
[hep-th/9206044]. 
\bibitem{Cadoni-2} M. Cadoni, Phys. Rev. {\bf D53}, 4413 (1996)
[gr-qc/9510012].
\bibitem{JT}  R. Jackiw in {\it Quantum Theory of Gravity}, ed. S. Christensen
(Hilger, Bristol,1984), p.403; C. Teitelboim, {\it ibid}, p.327;
R. Jackiw, Nucl. Phys. {\bf B252}, 343 (1985).
\bibitem{CGHS} C.G. Callan, S.B. Giddings, J.A. Harvey and A. Strominger,
Phys. Rev. {\bf D45}, 1005 (1992) [hep-th/9111056].
\bibitem{Cadoni-3} M. Cadoni and S. Mignemi, Phys. Lett. {\bf B358}, 217
(1995) [hep-th/9410041].
\bibitem{Hawking-3} G.W. Gibbons and S.W. Hawking, Phys. Rev. {\bf D15},
2752 (1977).
\bibitem{Wald} R.M. Wald, Phys. Rev. {\bf D48}, 3427 (1993) [gr-qc/9307038];
V. Iyer and R.M. Wald, Phys. Rev. {\bf D50}, 846 (1994) [gr-qc/9403028].
\bibitem{Birmingham-2} D. Birmingham, M. Blau, M. Rakowski and G. Thompson,
Phys. Rept. {\bf 209}, 129 (1991).
\bibitem{Joey-3} A.J.M. Medved, D. Martin and M. Visser, 
Class. Quant. Grav. {\bf 21}, 1393 (2004) [gr-qc/0310009].
\bibitem{Gabor-5} D. Louis-Martinez and G. Kunstatter, Phys. Rev. {\bf D52},
3494 (1995) [gr-qc/9503016].
\bibitem{Joey-2} A.J.M. Medved and G. Kunstatter, Phys. Rev. {\bf D59},
104005 (1999) [hep-th/9811052].
\bibitem{Krasnov} K. Krasnov and S.N. Solodukhin,
``Effective stringy description of Schwarzschild black holes'',
 arXiv:hep-th/0403046 (2004).
\bibitem{Conformal} V. de Alfaro, S. Fubini and G. Furlan, Nuovo.
Cim. {\bf 34A}, 569 (1976).
\bibitem{Vaidya} T.R. Govindarajan, V. Suneeta and S. Vaidya,
Nucl. Phys. {\bf B583}, 291 (2000) [hep-th/0002036].
\bibitem{Birmingham} D. Birmingham, K.S. Gupta and S. Sen,
Phys. Lett. {\bf B505}, 191 (2001) [hep-th/0102051].
\bibitem{Carlip} S. Carlip, Phys. Rev. Lett. {\bf 82}, 2828 (1999)
[hep-th/9812013]; Class. Quant. Grav. {\bf 16}, 3327 (1999)
[gr-qc/9906126]; Phys. Rev. Lett. {\bf 88}, 241301 [gr-qc/0203001].
\bibitem{Solodukhin} S.N. Solodukhin, Phys. Lett. {\bf B454},
213 (1999) [hep-th/9812056]. 
\bibitem{Bessel} {\it Handbook of Mathematical Functions}, ed. M. Abramowitz
and I.A. Stegun (Dover Publications,  New York, 1964).



\end{thebibliography}
\end{document}